\documentclass[mathleft
]{an}
\usepackage{graphicx}
\usepackage{times}
\usepackage{verbatim} 
\usepackage{natbib}
\bibpunct{(}{)}{;}{a}{}{,}
\def \nineteen {XB\,1916$-$053}
\def \mxb {MXB\,1658$-$298}
\def \bigdip {X\,1624$-$490}
\def \twelve {XB\,1254$-$690}
\def \grs {GRS\,1915+105}
\def \gro {GRO\,J1655$-$40}
\def \gx {GX\,13+1}
\def \cir {Cir\,X$-$1}
\def \exo {EXO\,0748$-$676}
\def \ax {AX\,J1745.6-2901}

\def \thirteen {4U\,1323$-$62}
\def \seventeen {H\,1743$-$322}
\def \threethreenine {GX\,339$-$4}
\def \xte {XTE\,J1650$-$500}
\def \sixteen {4U\,1630$-$47}
\newcommand \maxithirteen {MAXI\,J1305$-$704}

\def \nhgal {$N{\rm _H^{Gal}}$}

\newcommand {\logxi} {$\log\xi$}

\newcommand {\kms} {km~s$^{-1}$}

\def \fetfive {{Fe}\,{\small XXV}}
\def \fetsix {{Fe}\,{\small XXVI}}

\newcommand {\approxgt} {\mathrel{\hbox{\rlap{\lower.55ex \hbox {$\sim$}}
        \kern-.3em \raise.4ex \hbox{$>$}}}}
\newcommand {\approxlt} {\mathrel{\hbox{\rlap{\lower.55ex \hbox {$\sim$}}
        \kern-.3em \raise.4ex \hbox{$<$}}}}

\newcounter{mycounterg}
\newcommand {\nlt} {$^{\alph{mycounterg}}$ \addtocounter{mycounterg}{1}}

\newcounter{mycounterh}
\newcommand {\nlf} {~$^{\alph{mycounterh}}$ \addtocounter{mycounterh}{1}}

\newcommand \msun {\ifmmode$M$_{\odot}\else M$_{\odot}$\fi}
\newcommand \Msun {\ifmmode$M$_{\odot}\else M$_{\odot}$\fi}
\newcommand {\degree} {$^{\circ}$}
\newcommand {\DT}{D{\'{\i}}az Trigo}
\newcommand {\BC}{Ba{\l}uci{\'n}ska-Church}
\newcommand {\DAI}{D'A{\'{\i}}}

\begin{document}

% The following seven commands are intended for editorial usage and should be ignored by
% the author(s).
\Pagespan{1}{}% Document's page range. 
% If second parameter is left empty, the last page is computed automatically.
\Yearpublication{}%
\Yearsubmission{2015}%
\Month{}%   
\Volume{}%  
\Issue{}% 
% \DOI{This.is/not.aDOI}% 

\title{Accretion disc atmospheres and winds in low-mass X-ray binaries}

\author{M. D\'iaz Trigo\inst{1}\fnmsep\thanks{Corresponding author:
  \email{mdiaztri@eso.org}\newline}
\and  L. Boirin\inst{2}
}
\titlerunning{Disc atmospheres and winds in LMXBs}
\authorrunning{M. D\'iaz Trigo \& L. Boirin}
\institute{
ESO, Karl-Schwarzschild-Strasse 2, D-85748 Garching bei M\"unchen, Germany
\and 
Observatoire astronomique de Strasbourg, Universit\'e de Strasbourg, CNRS, UMR 7550, 11 rue de l'Universit\'e, F-67000 Strasbourg, France
}

\received{1 Sep 2015}
\accepted{}
\publonline{later}


  \abstract{In the last decade, X-ray spectroscopy has enabled a wealth
  of discoveries of photoionised absorbers in X-ray binaries. Studies
  of such accretion disc atmospheres and winds are of fundamental
  importance to understand accretion processes and possible feedback
  mechanisms to the environment. In this work, we review the current
  observational state and theoretical understanding of accretion disc
  atmospheres and winds in low-mass X-ray binaries, focusing on the
  wind launching mechanisms and on the dependence on accretion
  state. We conclude with issues that deserve particular attention.}

\maketitle

\section{Introduction}

The presence of highly ionised plasma in low-mass X-ray binaries
(LMXBs) was first observed in the microquasars \gro\
\citep{1655:ueda98apj} and \grs\ \citep{1915:kotani00apj} thanks to
the X-ray observatory {\it ASCA}. Their spectra showed narrow
absorption lines that were identified with absorption from \fetfive\
and \fetsix. Those first detections paved the way for a myriad of
discoveries of photoionised plasmas in LMXBs (see Table~\ref{obs})
following the launch of the X-ray observatories {\it Chandra},
XMM-Newton and {\it Suzaku}.

Soon after the first handful of detections was it realised that
photoionised absorbers could be ubiquitous to all X-ray binaries
\citep[e.g., ][]{1624:parmar02aa}.
However, only recently has it been recognized that photoionised plasmas may be a
key ingredient for our understanding of X-ray binaries (XRBs). The
reason is that the amount of mass that leaves the XRB when the plasmas
are outflowing is of the order of or larger than the mass accreted
into the system \citep[e.g., ][]{ponti12mnras} and could therefore
lead to instabilities in the accretion flow \citep[hereafter
BMS83]{begelman83apja} or even trigger accretion state changes
\citep{shields86apj}.

In this work, we review the current observational state and
theoretical understanding of accretion disc atmospheres and winds in
LMXBs.  High mass X-ray binaries are not discussed because their winds
are often dominated by the contribution from the companion star.
As a diagnostic tool for this study we focus on the presence of narrow
absorption lines in the spectra, with or without blueshifts. 

\section{Observational properties of accretion disc atmospheres and winds}

To date, 19 LMXBs have shown narrow absorption lines from highly
ionised ions situated in a ``local'' absorber inside the system as
opposed to the interstellar medium: 11 neutron star (NS) and 8 black
hole (BH) systems (Table~\ref{obs}).
The majority of these systems are dippers \citep[16/19 or 84\% in
Table~\ref{obs}, see also
][]{1916:boirin04aa,diaztrigo06aa,ponti12mnras}. These sources are
viewed relatively close to edge-on, i.e., at a high inclination angle
of 60--80\degree\ \citep{frank87aa}. The three systems not showing
dips in Table~\ref{obs} are suspected to have a relatively high,
$>$45-50$^{\circ}$, inclination.  The preferential detection of
absorbing plasmas at inclinations $\approxgt$50$^\circ$ points to a
distribution of the ionised plasma close to the accretion disc. This
suggests an equatorial geometry or a flared geometry exhibiting a
stratification of density and/or ionisation \citep{higginbottom15apj}.

With the aid of photoionised plasma models produced by {\tt XSTAR}
\citep{kallman01apjs} or {\tt CLOUDY} \citep{ferland93araa}, the
relative depths of the absorption lines detected in a spectrum give
access to the column density of the photoionised plasma and to its
degree of ionisation, defined as $\xi~=~L/nr^2$ \citep{tarter69apj},
where $L$ is the luminosity of the ionising source, $n$ the electron
density of the plasma and $r$ the distance between the plasma and the
ionising source. On the other hand, a blueshift of the lines with
respect to the theoretical wavelengths indicates that the plasma is
outflowing.

The ionised plasmas detected in LMXBs have column densities ranging
between 0.5\,$\times$10$^{21}$ and 10$^{24}$ cm$^{-2}$. There is no
obvious difference between the column densities observed for NSs and
BHs.
Among the sources for which \logxi\ was estimated, the vast majority
(16/18 or 89\% in Table~\ref{obs}) have shown very highly ionised
plasmas with \logxi$\geq$3. All these sources have indeed shown
absorption lines from \ion{Fe}{xxv} at 6.70~keV and/or from
\ion{Fe}{xxvi} at 6.97~keV.  Four sources have shown both high
(\logxi$\geq$3) and low (\logxi$<$3) ionisation plasmas.  Two sources
have only shown plasmas with \logxi$<$3: \maxithirteen\ and
\threethreenine. \maxithirteen\ has actually shown lines that may be
modeled by two absorbers, among which the most ionised one could still
have a \logxi\ close to 3 \citep[2.9 in ][]{1305:shidatsu13apj}.
\threethreenine\ is the only source left showing only one low
ionisation plasma \citep[\logxi\ = 1.8, ][]{gx339:miller04apj} with
no accompagning absorber detected with a higher degree of ionisation.
Interestingly, \threethreenine\ does not show dips, suggesting that
there could be stratification of the ionised plasma as a function of
inclination. 
On the other hand,
most of the low ionisation plasmas have been detected in systems that
show a relatively low column density of interstellar absorption,
indicating that the number of systems with low ionisation plasmas may
be actually larger.

We note that as deeper observations become available, the need to
include more than one ionised plasma while modelling a given spectrum
is increasing due to the co-existence of lines with significantly
different outflow velocities or different ionisation equilibria
\citep[e.g., ][]{1624:xiang09apj,1655:kallman09apj}. This suggests
that studies of absorption measurement distributions as done for AGN
\citep[e.g., ][]{stern14mnras} could soon be needed also in XRBs.

We consider that a given system shows an outflow if the absorption
lines show a significant blueshift or a P-Cygni profile. Conversely,
we call atmospheres those plasmas for which no blueshift is detected
with the high resolution gratings on board {\it Chandra} and
XMM-Newton. Typically, these gratings enable to derive upper-limits on
velocity shifts of $\sim$250~\kms\ \citep[e.g., ][]{1916:juett06apj}
while CCD instruments like the pn camera on board XMM-NEWTON give
upper-limits of $\sim$1000~\kms\ \citep[e.g., ][]{1323:boirin05aa}.
Roughly half of the sources in Table~\ref{obs} show an outflow (8
among the 15 systems showing lines clearly attributed to a local
absorber and whose wavelength was measured with gratings). This
includes all the BH LMXBs showing dips and the BH IGR~J17091--3624 not
showing dips.  In contrast, a significant blueshift in the absorption
lines could only be measured for $\sim$~30\% (3/11) of the NS LMXBs
showing an ionised absorber. The overwhelming presence of outflows in
BHs as opposed to NSs could be mostly due to the size of the systems
since the NSs showing an outflow are precisely those with large
orbital periods of the order of those found in BHs. Line blueshifts
are moderate, between $\sim$\,400 and 3000~km s$^{-1}$, for the
majority of systems that show an outflow.
There is an indication for an ultra fast outflow in IGR~J17091--3624
\citep[a single line detected at 4$\sigma$ significance blueshifted by
$\sim$9300~\kms, not found in a second observation of the same source
during a similar soft state,][]{1709:king12apj}.
In \maxithirteen, \cite{1305:miller14apj} reported the detection of a
rich absorption complex in the Fe-L band redshifted by 540 to
14390~\kms, indicating an infall of the photoionised plasma that may
correspond to a failed wind.
  
One important difference between systems showing static atmospheres
and outflows is that only the latter will undergo mass loss.
Mass-loss rates are of the order of 1--20 times the mass accretion
rates in the black hole systems \citep[e.g., ][]{1915:lee02apj,1915:neilsen11apj,1709:king12apj,ponti12mnras}, implying
that the wind presence or absence along the outburst (see Sect.~4) could be key to understand the outburst evolution 
\citep{shields86apj}.

\begin{table*}
  
  \caption{List of low-mass X-ray binaries for which a photoionised plasma local to the source has been detected in absorption. The first columns indicate the source name, the orbital period, the galactic column density towards the source after \cite{kalberla05aa}, whether the compact object is a neutron star (NS) or a black hole, whether it shows dips (D), and, if not, the source inclination ($i$). See \cite{liu07aa} for the references on the period and inclination, unless otherwhise noted. The last columns indicate if the degree of ionization of the photoionised  absorber(s) detected during persistent intervals (outside dips) is $< 3$ or $\geq 3$ in \logxi, whether the absorber seems to  flow outwards (out), inwards (in) or to be bound as an atmosphere (atm), based on the velocity shifts of the absorption lines detected with {\it Chandra} HETGS or XMM-Newton RGS. ``no grat.'' means that no contraints were published on the velocity shifts from these gratings.}
\label{obs}

\begin{tabular}{l@{\extracolsep{1mm}}c@{\extracolsep{-1mm}}c@{\extracolsep{0mm}}c@{\extracolsep{1mm}}c@{\extracolsep{0mm}}c@{\extracolsep{0mm}}c@{\extracolsep{0mm}}c@{\extracolsep{0mm}}c@{\extracolsep{1mm}}p{0.45\textwidth}}
\hline
\noalign{\smallskip}
Source & P$_{\rm orb}$ & \nhgal             & NS & Dips & $i$ ($^\circ$) & \multicolumn{2}{c}{\logxi} & Flow & References on the warm absorbers\\ 
       &             & 10$^{21}$~cm$^{-2}$  &    &    &   &    $< 3$ & $\geq 3$         &      &            \\
\noalign{\smallskip}
\hline
\noalign{\smallskip}
\nineteen  & 	0.83 h  & 	2.3  & 	NS & 	D & 	 & 	x & 	x & 	atm & 	Boirin04, Juett06, \DT06, Iaria06, Zhang14\\
1A 1744--361  & 	1.62 h  & 	3.1  & 	NS & 	D & 	 & & 	x & 	atm & 	Gavriil12\\
\thirteen  & 	2.93 h  & 	12 & 	NS & 	D & 	 & 	 & 	x & 	no grat. & Boirin05, Church05, \BC09\\
\exo\  & 	3.82 h  & 	1.0  & 	NS & 	D & 	 & 	x & 	x & 	atm & 	\DT06, van Peet09, Ponti14\\
\twelve  & 	3.93 h  & 	2.0  & 	NS & 	D & 	 & 	 & 	x & 	atm & 	Boirin03, \DT06/09 , Iaria07\\
\mxb  & 	7.11 h  & 	1.9  & 	NS & 	D & 	 & 	x & 	x & 	atm & 	Sidoli01, \DT06\\
\xte\  & 	7.63 h  & 	4.2  & 	 & 	 & $>$ 50 & 	?\nlt & ?\nlt & ?\nlt & 	Miller02/04 \\
\ax\  & 	8.4 h  & 	12 & 	NS & 	D & 	 & 	 & 	x & 	no grat. & 	Hyodo09, Ponti15 \\
MAXI J1305--704  & 9.74 h\nlt & 	1.9  & 	 & 	D & 	 & 	x & 	 & in & 	Shidatsu13, Miller14\\
\bigdip  & 	20.89 h  & 	20 & 	NS & 	D & 	 & 	 & 	x & 	atm & 	Parmar02, \DT06, Iaria07b, Xiang09\\
IGR J17480--2446  & 21.27 h\nlt & 	6.5  & 	NS & 	D & 	 & & 	x & 	out & 	Miller11\\
\threethreenine\  & 1.76 d  & 	3.6  & 	 & 	 & $>$ 45\nlt & x & 	 & 	?\nlt  & 	Miller04, Juett06\\
\gro\  & 	2.62 d  & 	5.2  & 	 & 	D & 	 & 	 & 	x & 	out & 	Ueda98, Yamaoka01, Miller06b/08, Netzer06, Sala07, \DT07, Kallman09, Luketic10, Neilsen12\\
\cir  & 	16.6 d  & 	16 & 	NS & 	D & 	 & 	x & 	x & 	out & 	Brandt00, Schulz02, , \DAI07, Iaria08, Schulz08\\
\gx  & 	24.06 d  & 	13 & 	NS & 	D & 	 & 	 & 	x & 	out & 	Ueda01/04, Sidoli02, \DT12, Madej14, D'Ai14\\
\grs  & 	33.5 d  & 	13 & 	 & 	D & 	 & 	 & 	x & 	out & 	Kotani00, Lee02, Martocchia06, Ueda09/10, Neilsen09/11/12\\
IGR J17091--3624  & 	$>$4 d\nlt & 	5.4  & 	 & 	 & $>$ 53\nlt & & x & 	out & 	King12\\
\sixteen\  & 	 & 	17 & 	 & 	D & 	 & 	 &     	    x & 	out & 	Kubota07, \DT13/14, King13/14, Neilsen14 \\
\seventeen  & 	 & 	6.9  & 	 & 	D & 	 & 	 &      	x & 	out & 	Miller06a\\
\noalign{\smallskip}
\hline
\noalign{\smallskip}
\multicolumn{10}{p{0.96\textwidth}}{%
  \nlf~not estimated; absorption lines from \ion{Ne}{ix} and of a \ion{Ne}{ii} \citep{gx339:miller04apj}.
  \nlf~not estimated; absorption feature near 7~keV possibly due to \ion{Fe}{xxvi} \citep{1650:miller02apjl}.
  \nlf~Detection of an unshifted line from \ion{Ne}{ix} and of a \ion{Ne}{ii} line blueshifted by $\sim510\pm60$~\kms\ possibly due to a local absorber \citep{gx339:miller04apj}. An interstellar origin was attributed to a similar \ion{Ne}{ii} line in  \threethreenine\ (see below).
\nlf~\cite{1305:shidatsu13apj}
\nlf~\cite{igrj1748:papitto11aa} % orbital period and dipping-like structures
\nlf~using the lower limit on the mass of the companion star estimated by  \cite{gx339:munoz-darias08mnras} and assuming that the
black hole mass is less than $15~\msun$ \citep{gx339:shidatsu11pasj}.
\nlf~Detection of several lines (\ion{Ne}{ix}, \ion{O}{vii}, etc) with blueshifts in the range 50--160~\kms\ and of lines from \ion{Ne}{ii-iii} blueshifted by $510\pm20$~\kms\ \citep{gx339:miller04apj}. While the \ion{Ne}{ix} line is produced  mainly by a local absorber, the \ion{Ne}{ii-iii} lines are consistent with being produced by the hot interstellar medium \citep{juett06apj}.
\nlf~\cite{1709:wijnands12mnras}
\nlf~\cite{1709:rao12apjl}.  
}\\
%\hline
\end{tabular}

\end{table*}

\nocite{1630:kubota07pasj,king13apj,1630:diaztrigo13nat,1630:neilsen14apjl,1630:king14apjl,1630:diaztrigo14aa}
\nocite{1743:miller06apj}
\nocite{1916:boirin04aa,1916:juett06apj,1916:iaria06apj,diaztrigo06aa,1916:zhang14pasj}
\nocite{1a1744:gavriil12apj}
\nocite{1323:church05mnras,1323:boirin05aa,1323:balucinska09aa}
\nocite{diaztrigo06aa,0748:vanpeet09aa,0748:ponti14mnras}
\nocite{1254:boirin03aa,1254:diaztrigo09aa,diaztrigo06aa,1254:iaria07aa}
\nocite{1658:sidoli01aa,diaztrigo06aa}
\nocite{1650:miller02apjl,gx339:miller04apj}
\nocite{axj1745:hyodo09pasj,axj1745:ponti15mnras}
\nocite{1305:shidatsu13apj,1305:miller14apj}
\nocite{1624:parmar02aa,diaztrigo06aa,1624:iaria07aa,1624:xiang09apj}
\nocite{igrj1748:miller11apj}
\nocite{gx339:miller04apj,juett06apj}
\nocite{1655:ueda98apj,1655:yamaoka01pasj,1655:miller06nat,1655:miller08apj,1655:kallman09apj,1655:luketic10apj,1655:netzer06apj,1655:neilsen12apj,1655:sala07aa,1655:diaztrigo07aa}
\nocite{1709:king12apj}
\nocite{cirx1:brandt00apjl,cirx1:schulz02apj,cirx1:iaria08apj,cirx1:schulz08apj,cirx1:dai07apj}
\nocite{gx13:ueda01apjl,gx13:sidoli02aa,gx13:ueda04apj,gx13:diaztrigo12aa,madej14mnras,gx13:dai14aa}
\nocite{1915:kotani00apj,1915:martocchia06aa,1915:lee02apj,1915:ueda09apj,1915:neilsen09nat,1915:ueda10apj,1915:neilsen11apj,1915:neilsen12mnras}

\section{The wind launching mechanism}
\label{launching}

The wealth of data collected on photoionised plasmas in LMXBs (see
Table~1) allows us to investigate the launching mechanism of winds in
these sources. The successful mechanism should be able to reproduce
the observed plasma characteristics (density, column density, outflow
velocity or degree of ionisation) for single observations from the
same source or from different sources. In addition, if there is a
universal mechanism for the formation of winds in LMXBs, it should be
able to explain the properties of the sample, i.e. why some systems
show static atmospheres and others outflows and why the plasmas are
preferentially observed at high inclinations or ``soft'' accretion
states (see Sect.~4).

Accretion disc winds can be driven via thermal, radiative and/or
magnetic pressure. In what follows we examine the possibility that
winds in LMXBs are launched via thermal pressure. However, it should
be kept in mind that the dominant mechanism could change as a system
evolves through an accretion outburst. For example, \citet{proga02apj}
demonstrated that, in general, radiation pressure due to lines cannot drive a wind in LMXBs
due to the strong irradiation of the ``UV-emitting disk" but radiation pressure due to
electron scattering assists thermal expansion to drive a hot wind if the luminosity 
rises above the Eddington limit.

Thermal pressure or Compton-heated winds are expected to arise in
systems in which the accretion disc is subject to X-ray irradiation
from the central part such as X-ray binaries or quasars (BMS83). In
these systems, the X-rays illuminating the disc can heat the gas to
temperatures exceeding $\sim$\,10$^7$~K predominantly through the
Compton process. The heated gas will then form an atmosphere or corona
above the disc \citep[BMS83; ][]{shakura73aa,jimenez02apj}. The upper
boundary of the atmosphere is the Compton temperature corona, which is
less dense and hotter than the underlying atmosphere. The evaporated
photoionised plasma will remain bound to the disc as an
atmosphere/corona or be emitted as a thermal wind, depending on
whether the thermal velocity exceeds the local escape velocity
\citep[BMS83;][]{woods96apj, proga02apj}. Importantly, the radial
extent of the corona is determined only by the mass of the compact
object and the Compton temperature and is independent of luminosity
\citep[BMS83;][]{woods96apj}. BMS83 determined that a wind would be
launched by thermal pressure at radii larger than 0.1\,r$_{IC}$ (where
r$_{IC}$ denotes the Compton radius or distance at which the escape
velocity equals the isothermal sound speed at the Compton temperature
T$_{IC}$). \citet{proga02apj} found that for very luminous systems,
including the radiation force due to electrons could lower the
effective gravity and subsequently the escape velocity and allowed a
hot robust disc wind to be already produced at $\sim$\,0.01\,$r_{IC}$,
well inside the Compton radius and previous estimates by BMS83 and
\citet{woods96apj}.

It is already apparent from Table~1 that outflows are only present in
systems with ``long'' orbital periods. This could indicate that the
capability to launch a wind is related to the size of the disc,
consistent with the expectations of thermal winds. The disc size can
be approximated as r$_{disc}\,\sim$\,0.8r$_L$, where r$_L$ is the
radius of the Roche lobe for the compact object and can be estimated
from the orbital period, the mass of the compact object and the mass
ratio following \citet{eggleton83apj} and Kepler's third law
\citep[see equations 4.2 and 4.6 from][]{frank02apia}.  We list in
Table~2 the disc sizes for all the sources from Table~1 for which
there are reliable estimates of the mass ratio and (for BHs) of the
mass of the compact object. We then compare these values to the radius
at which Compton-heated winds could be launched, $\sim$ 0.1 r$_{IC}$
(see above). It follows that small systems like \nineteen\ will be
incapable of launching winds via the thermal mechanism.  
In contrast, large systems like \gro\ and \grs\ will be able to launch
thermal winds (regardless of whether additional mechanisms are in
play).
If the Compton temperature is well above 10$^{7}$~K, the Compton
radius would proportionally decrease allowing the launch of winds in
smaller systems. However, T$_{IC}$ $\sim$10$^{7}$~K seems a fair
assumption given that winds are preferentially launched during
``soft'' states (see Sect. 4).

In conclusion, Table~2 shows that at the extremes of small and large
disc sizes, the expectations of thermal winds are fulfilled in the
sense that small systems show atmospheres and large systems show winds
\citep[see also Fig.~1 from \citet{diaztrigo13actapoly} for a
comparison of the estimated position of the plasma and source
luminosity for NS LMXBs with the expectations of thermal winds
from][]{woods96apj}. In contrast, for systems of medium sizes this
test is inconclusive since sources like \twelve\ or \mxb\ could in
principle show mild thermal winds but only static atmospheres have
been reported.

\begin{table*}
\centering%%%
\caption{LMXBs for which a photoionised absorber local to the source has been detected and for which there are good estimates for the period, the 
mass ratio $q$ and (for BHs) the mass of the compact object (M$_1$). We include NSs for which their mass has not been estimated if the period and
the mass ratio are known and fix the NS mass to 1.4$\Msun$. We estimate $r_{IC}$ = 10$^{10}$ T$_{IC8}^{-1}$ (M$_1$/$\Msun$) cm (eq. 2.7 from BMS83) and assume T$_{IC8}$ = 0.1 (or 10$^7$ K).}
\label{obs2}
\begin{tabular}{lcccccc}\hline
Source & Period & M$_1$ &  q & Reference & 0.8 r$_L$ & 0.1 r$_{IC}$ \\ 
& [hrs] & ($\Msun$) & & & [10$^{10}$ cm]  &  [10$^{10}$ cm]  \\
\hline
\nineteen & 0.83 & 1.2--1.6 & 0.046  & \citet{1916:heinke13apj} & 0.43--0.47 & 1.2--1.6 \\
\exo\ & 3.82 & 1.0--2.4 & 0.11--0.28 & \citet{0748:munoz09mnras} & 1.5--2.7 & 1.0--2.4 \\
\twelve & 3.93 & 1.2--1.8 & 0.33--0.36 & \citet{1254:cornelisse13mnras} & 2.3--2.8 & 1.2--1.8 \\
\mxb & 7.11 &  1.4 & 0.18--0.64\nlt & \citet{1658:cominsky84apj} & 3.0--4.7 & 1.4 \\
IGR J17480--2446 & 21.27 &  1.4  & 0.3--1.1\nlt & \citet{igrj1748:papitto11aa} & 7.4--12.0 & 1.4 \\
\gro\ & 62.92 & 6.6\,$\pm$\,0.5 & 0.42\,$\pm$\,0.03 & \citet{casares14ssr} & 27.3--30.2 & 6.1--7.1 \\  
\grs & 812.4 & 10.1\,$\pm$\,0.6  & 0.042\,$\pm$\,0.024 &\citet{casares14ssr} & 69.6--100.5 & 9.5--10.7 \\
\noalign{\smallskip}
\hline
\noalign{\smallskip}
\multicolumn{7}{p{0.92\textwidth}}{%
\nlf The mass ratio has been estimated assuming a NS of 1.4$\Msun$ and the companion mass estimated by \citet{1658:cominsky84apj} of 0.25--0.9$\Msun$.
\nlf The mass ratio has been estimated assuming a NS of 1.4$\Msun$ and the companion mass estimated by \citet{igrj1748:papitto11aa} of 0.42--1.5$\Msun$.
  }
\end{tabular}
\end{table*} 
 
However, regardless of whether a system could launch
Compton-heated winds due to its size, we should actually prove that
the location of the plasma that is outflowing is outside the
Compton radius for that system. Unfortunately, this is currently challenging
for the majority of the observed winds due (mainly) to the uncertainty
in the density of the plasma. As an example, a plasma with log$\xi \sim$ 3
subject to a ionising luminosity of 10$^{36}$ erg s$^{-1}$ would be
located at 10$^{10}$ cm for a plasma density of 10$^{13}$ cm$^{-3}$
and therefore be able to outflow (using $\xi~=~L/nr^2$, see Sect.~2). 
For a larger density of 10$^{15}$
cm$^{-3}$ the plasma would be instead located at 10$^{9}$ cm and
therefore bound to the system. To date the direct measurements of
plasma density are rare \citep[for two sources densities of 10$^{13}$--10$^{15}$~cm$^{-3}$ have been determined,][]{cirx1:schulz08apj,1655:kallman09apj}. Based on the density
measured for \gro, \citet{1655:kallman09apj} derived a radius of
7\,$\times$\,10$^9$~cm for the location of the plasma and concluded
that the wind could not be thermally launched, in agreement with
previous claims of a magnetic nature for this wind \citep[][but see \citet{1655:netzer06apj}]{1655:miller06nat,1655:miller08apj}. \citet{1655:luketic10apj} performed
hydrodynamical simulations and also concluded that the wind could not be
thermally launched, since they could not find a space in their
simulations which showed simultaneously a high density ($>$10$^{12}$ cm$^{-3}$) and velocity
outflow, as detected in the {\it Chandra} observation (but note that it remains to be proven 
that magnetic pressure can generate such a dense, fast wind, see e.g. Chakravorty et al., this proceedings). To date this is
the only claim of a disc wind driven by magnetic pressure and is
therefore important to understand its launching conditions and the differences with respect to other epochs of the same system or different systems. \citet{diaztrigo13actapoly} pointed out that the luminosity
of the system could have been severely underestimated for the measured
plasma column densities \citep[$\sim$10$^{24}$ cm$^{-2}$,][]{1655:miller08apj} by
not including Compton scattering in spectral modelling.  
The fact that the optical flux was
increasing monotonically around the epoch of the {\it Chandra}
observation, while the X-ray flux started decreasing 10 days before
suggests that the wind was optically thick to Compton scattering and
that the X-rays were scattered and absorbed by the wind, giving
further support to the idea that the luminosity was
underestimated (Shidatsu et al., these proceedings). Interestingly,
\citet{higginbottom15apj} succeeded in simulating thermal winds
that are faster and denser than found by \citet{1655:luketic10apj} by
increasing the ionising luminosity or reducing line cooling while increasing X-ray heating. Clearly, the issue of wind opacity
and potential underestimation of luminosity deserves further studies
before we can discard a thermal launching mechanism for this
wind. Different Compton temperatures, obtained by fitting plausible
continuum models to broadband spectra, should also be considered when
calculating the allowed radii since the non-modelling of Compton
scattering could introduce significant changes in the Spectral Energy Distribution (SED).

In summary, the potentially magnetically
driven wind from \gro\ should be studied further due to its
uniqueness. However, given that all other sources and even other stages of the wind from \gro\
could be explained as generated by thermal pressure \citep[e.g.][]{1655:neilsen12apj}, 
we conclude that Compton heating is an excellent candidate as the dominant launching mechanism
of winds in LMXBs. This is also supported by the fact that both winds and
atmospheres are only detected in high-inclination sources
(see Table~1), consistent with the geometry predicted for a
thermal wind that will be preferentially observed close to the
equatorial plane, since at polar angles, $\approxlt$\,45\,$^{\circ}$,
the low density and the high ionisation of the gas prevent its
detection.

\section{Dependence on accretion state}

To date, disc winds have been predominantly observed in the
``high/soft'', thermal-dominated, state of BH transients, when the jet
emission is absent. Conversely, in observations of the ``low/hard''
state of the same transients, with typical jet emission, winds were
excluded. In particular, blueshifted absorption lines were observed in
the soft state spectra of \gro\ \citep{1655:miller06nat,
  1655:diaztrigo07aa}, \grs\ \citep{1915:ueda09apj}, \sixteen\
\citep{1630:kubota07pasj, 1630:diaztrigo14aa} and \seventeen\
\citep{1743:miller06apj}. In contrast, observations of the hard state
of \gro\ \citep{1655:takahashi08pasj} and \seventeen\
\citep{1743:miller12apjl} excluded the presence of such lines down to
equivalent widths of 20 and 3 eV, respectively. For \grs, disc winds
are predominantly observed during soft states (Neilsen \& Lee 2009,
Ueda et al. 2009) but there is also one detection of a weak wind in a
``hard'' (or ``C'') state \citep{1915:lee02apj}. The winds appear to
be absent also in the ``very high state'' (VHS) of \sixteen\
\citep{1630:diaztrigo14aa} and \seventeen\
\citep{1743:miller06apj}. The presence of \fetfive\ and \fetsix\ lines
as a function of accretion state has been recently investigated for
two NS XRBs, \exo\ and \ax, by
\citet{0748:ponti14mnras,axj1745:ponti15mnras} who concluded that the
dichotomy observed for BH transients was also present for NSs.

At this stage it is important to establish if the absence
of atmospheres/winds can be derived from the absence of absorption
from \fetfive\ or \fetsix. 
While there is only one case where absorption from \fetsix\ has been
found during a hard state \citep[][see above]{1915:lee02apj}, the
presence of a photoionised plasma during hard states has been inferred
from a broadened Pa$\beta$ emission line in \threethreenine\
\citep{gx339:rahoui14mnras} and from absorption troughs below 2~keV in
\exo\ \citep{diaztrigo06aa,0748:vanpeet09aa}. The plasma had a low
degree of ionisation in both cases.
Therefore, it might be that atmospheres/winds are present during
states other than the soft but their degree of ionisation
or density is such that no lines are expected from highly ionised
iron.

Clearly, if photoionised plasmas in LMXBs are
the result of disc irradiation, then we expect important
changes in such plasmas as the systems transit from a soft to a hard state due to the change of the SED, which determines the Compton temperature (see
Sect.~3). However, it remains to be determined if the change of SED is
the decisive factor for the existence of winds or if other components
such as the jet play a significant role.  Based on \grs\ observations,
\citet{1915:neilsen09nat} proposed that 
the wind
observed during the soft state carries enough mass away from the disc
to halt the flow of matter into the radio
jet. \citet{chakravorty13mnras} invoked instead the thermal
instability present during the hard state of XRBs as a plausible
inhibitor of winds in that state. \citet{1630:diaztrigo14aa} followed
the evolution of the wind in \sixteen\ 
as the luminosity increased along the soft state and the transition to a VHS. They observed a decrease in the column density and
an increase of ionisation of the wind as the SED became harder towards
higher luminosities and concluded that the disappearance of the wind
in the VHS was a consequence of over-ionisation (note that {\it both} the increase of Compton
temperature, due to the hardening of the SED, and luminosity play a role). In contrast,
\citet{1630:hori14apj} compared observations of \sixteen\ from two
 outbursts and argued that ionisation
was important but not enough to make the wind disappear during the VHS. Other
authors have pointed to the possibility of density
\citep{1630:hori14apj} or geometrical
\citep{1743:miller06apj,1915:ueda10apj} changes to explain the
presence of atmospheres/winds in some accretion states and their
absence in others.

Most likely several of the effects above play a role in producing an
``observable'' atmosphere/wind. For example, a change in
luminosity could induce a change in the plasma density and
leave its level of ionisation unchanged \citep{higginbottom15apj}.
Conversely, a hardening/softening of the
spectrum implies a change of Compton temperature and consequently of
the Compton radius. 
This could translate into a different wind formation region.
In contrast, it is not clear that the presence of a thermal
instability should inhibit the wind since the presence of ionised
plasma is detected in simulations where gas was thermal unstable 
\citep{higginbottom15apj}. In this context it is important
to derive reliable SEDs including Compton
scattering to infer the state of the plasma expected at
different accretion states. For example, the discrepant conclusion
about over-ionisation causing the disappearance of the wind in the VHS
of \sixteen\ (see above) could be a result of using different SEDs. A
realistic SED reconstruction might explain not
only why winds are preferentially detected in some accretion states
but also why a wind may disappear during a given accretion state
\citep{1709:king12apj}.

\section{Conclusions and prospects}

There are excellent prospects for advancement in this field in the
coming years. In the observational domain the launch of Astro-H,
foreseen for 2016, will allow fine spectroscopy of photoionised
plasmas with its X-ray Calorimeter Spectrometer and reliable
determination of broad-band SEDs thanks to its hard X-ray Imaging
System. In the theoretical and modelling domain, two issues deserve
immediate attention to progress further. The first item is the
inclusion of Compton scattering both in hydrodynamical simulations and modelling
of photoionised plasmas since this determines how well we can infer
the incident SED and luminosity that irradiates the disc. Moreover,
the correlation between the evolution of the wind and the broad iron
line observed in two LMXBs
\citep{gx13:diaztrigo12aa,1630:diaztrigo14aa} indicates that Compton
scattering in the wind could actually account for a fraction of the
broad iron lines that are observed in XRBs, having consequences for
the measurement of black hole spin with methods that request the
broad iron line to purely arise from disc reflection. The second issue
is related to the (most likely) over-simplistic constant density
models for photoionised plasmas that are used in the vast majority of
the literature. While so far most of the sources included in Table~1
show one or at most two states of the plasma simultaneously, it is
expected that this will change as the observations become more
sensitive. In this context, it may be important to consider plasmas of
different densities co-existing in the same state of a source (as it
is the case for plasmas under constant pressure conditions) or a
change in density between different accretion states.

Considering the above and bearing in mind that XRB winds could be the
key missing ingredient to understand accretion state changes during an
outburst we are expectant for discoveries in the next decade.

\acknowledgements
We thank the anonymous referee for 
helpful comments. We also thank Daniel Proga for a careful reading of this manuscript and the suggestions
that helped to improve it.

\bibliography{mybib}   % name your BibTeX data base

\begin{thebibliography}{96}
\providecommand{\natexlab}[1]{#1}

\bibitem[{{Ba{\l}uci{\'n}ska-Church} et~al.(2009){Ba{\l}uci{\'n}ska-Church},
  {Dotani} et~al.}]{1323:balucinska09aa}
{Ba{\l}uci{\'n}ska-Church}, M., {Dotani}, T. et~al.: 2009 \aap 500, 873

\bibitem[{{Begelman} et~al.(1983){Begelman}, {McKee} et~al.}]{begelman83apja}
{Begelman}, M.~C., {McKee}, C.~F. et~al.: 1983 \apj 271, 70

\bibitem[{{Boirin} et~al.(2005){Boirin}, {M{\' e}ndez}
  et~al.}]{1323:boirin05aa}
{Boirin}, L., {M{\' e}ndez}, M. et~al.: 2005 \aap 436, 195

\bibitem[{{Boirin} \& {Parmar}(2003)}]{1254:boirin03aa}
{Boirin}, L. \& {Parmar}, A.~N.: 2003 \aap 407, 1079

\bibitem[{{Boirin} et~al.(2004){Boirin}, {Parmar} et~al.}]{1916:boirin04aa}
{Boirin}, L., {Parmar}, A.~N. et~al.: 2004 \aap 418, 1061

\bibitem[{{Brandt} \& {Schulz}(2000)}]{cirx1:brandt00apjl}
{Brandt}, W.~N. \& {Schulz}, N.~S.: 2000 \apjl 544, L123

\bibitem[{{Casares} \& {Jonker}(2014)}]{casares14ssr}
{Casares}, J. \& {Jonker}, P.~G.: 2014 \ssr 183, 223

\bibitem[{{Chakravorty} et~al.(2013){Chakravorty}, {Lee}
  et~al.}]{chakravorty13mnras}
{Chakravorty}, S., {Lee}, J.~C. et~al.: 2013 \mnras 436, 560

\bibitem[{{Church} et~al.(2005){Church}, {Reed} et~al.}]{1323:church05mnras}
{Church}, M.~J., {Reed}, D. et~al.: 2005 \mnras 359, 1336

\bibitem[{{Cominsky} \& {Wood}(1984)}]{1658:cominsky84apj}
{Cominsky}, L.~R. \& {Wood}, K.~S.: 1984 \apj 283, 765

\bibitem[{{Cornelisse} et~al.(2013){Cornelisse}, {Kotze}
  et~al.}]{1254:cornelisse13mnras}
{Cornelisse}, R., {Kotze}, M.~M. et~al.: 2013 \mnras 436, 910

\bibitem[{{D'A{\'{\i}}} et~al.(2007){D'A{\'{\i}}}, {Iaria}
  et~al.}]{cirx1:dai07apj}
{D'A{\'{\i}}}, A., {Iaria}, R. et~al.: 2007 \apj 671, 2006

\bibitem[{{D'A{\`i}} et~al.(2014){D'A{\`i}}, {Iaria} et~al.}]{gx13:dai14aa}
{D'A{\`i}}, A., {Iaria}, R. et~al.: 2014 \aap 564, A62

\bibitem[{{D{\'{\i}}az Trigo} \& {Boirin}(2013)}]{diaztrigo13actapoly}
{D{\'{\i}}az Trigo}, M. \& {Boirin}, L.: 2013 Acta Polytechnica 53, 659

\bibitem[{{D{\'{\i}}az Trigo} et~al.(2014){D{\'{\i}}az Trigo}, {Migliari}
  et~al.}]{1630:diaztrigo14aa}
{D{\'{\i}}az Trigo}, M., {Migliari}, S. et~al.: 2014 \aap 571, A76

\bibitem[{{D{\'{\i}}az Trigo} et~al.(2013){D{\'{\i}}az Trigo}, {Miller-Jones}
  et~al.}]{1630:diaztrigo13nat}
{D{\'{\i}}az Trigo}, M., {Miller-Jones}, J.~C.~A. et~al.: 2013 \nat 504, 260

\bibitem[{{D{\'{\i}}az Trigo} et~al.(2006){D{\'{\i}}az Trigo}, {Parmar}
  et~al.}]{diaztrigo06aa}
{D{\'{\i}}az Trigo}, M., {Parmar}, A.~N. et~al.: 2006 \aap 445, 179

\bibitem[{{D{\'{\i}}az Trigo} et~al.(2007){D{\'{\i}}az Trigo}, {Parmar}
  et~al.}]{1655:diaztrigo07aa}
{D{\'{\i}}az Trigo}, M., {Parmar}, A.~N. et~al.: 2007 \aap 462, 657

\bibitem[{{D{\'{\i}}az Trigo} et~al.(2009){D{\'{\i}}az Trigo}, {Parmar}
  et~al.}]{1254:diaztrigo09aa}
{D{\'{\i}}az Trigo}, M., {Parmar}, A.~N. et~al.: 2009 \aap 493, 145

\bibitem[{{D{\'{\i}}az Trigo} et~al.(2012){D{\'{\i}}az Trigo}, {Sidoli}
  et~al.}]{gx13:diaztrigo12aa}
{D{\'{\i}}az Trigo}, M., {Sidoli}, L. et~al.: 2012 \aap 543, A50

\bibitem[{{Eggleton}(1983)}]{eggleton83apj}
{Eggleton}, P.~P.: 1983 \apj 268, 368

\bibitem[{{Ferland}(2003)}]{ferland93araa}
{Ferland}, G.~J.: 2003 \araa 41, 517

\bibitem[{{Frank} et~al.(2002){Frank}, {King} et~al.}]{frank02apia}
{Frank}, J., {King}, A. et~al.: 2002 \emph{{Accretion Power in Astrophysics:
  Third Edition}}

\bibitem[{{Frank} et~al.(1987){Frank}, {King} et~al.}]{frank87aa}
{Frank}, J., {King}, A.~R. et~al.: 1987 \aap 178, 137

\bibitem[{{Gavriil} et~al.(2012){Gavriil}, {Strohmayer}
  et~al.}]{1a1744:gavriil12apj}
{Gavriil}, F.~P., {Strohmayer}, T.~E. et~al.: 2012 \apj 753, 2

\bibitem[{{Heinke} et~al.(2013){Heinke}, {Ivanova} et~al.}]{1916:heinke13apj}
{Heinke}, C.~O., {Ivanova}, N. et~al.: 2013 \apj 768, 184

\bibitem[{{Higginbottom} \& {Proga}(2015)}]{higginbottom15apj}
{Higginbottom}, N. \& {Proga}, D.: 2015 \apj 807, 107

\bibitem[{{Hori} et~al.(2014){Hori}, {Ueda} et~al.}]{1630:hori14apj}
{Hori}, T., {Ueda}, Y. et~al.: 2014 \apj 790, 20

\bibitem[{{Hyodo} et~al.(2009){Hyodo}, {Ueda} et~al.}]{axj1745:hyodo09pasj}
{Hyodo}, Y., {Ueda}, Y. et~al.: 2009 \pasj 61, 99

\bibitem[{{Iaria} et~al.(2008){Iaria}, {D'A{\'{\i}}} et~al.}]{cirx1:iaria08apj}
{Iaria}, R., {D'A{\'{\i}}}, A. et~al.: 2008 \apj 673, 1033

\bibitem[{{Iaria} et~al.(2006){Iaria}, {Di Salvo} et~al.}]{1916:iaria06apj}
{Iaria}, R., {Di Salvo}, T. et~al.: 2006 \apj 647, 1341

\bibitem[{{Iaria} et~al.(2007{\natexlab{a}}){Iaria}, {di Salvo}
  et~al.}]{1254:iaria07aa}
{Iaria}, R., {di Salvo}, T. et~al.: 2007{\natexlab{a}} \aap 464, 291

\bibitem[{{Iaria} et~al.(2007{\natexlab{b}}){Iaria}, {Lavagetto}
  et~al.}]{1624:iaria07aa}
{Iaria}, R., {Lavagetto}, G. et~al.: 2007{\natexlab{b}} \aap 463, 289

\bibitem[{{Jimenez-Garate} et~al.(2002){Jimenez-Garate}, {Raymond}
  et~al.}]{jimenez02apj}
{Jimenez-Garate}, M.~A., {Raymond}, J.~C. et~al.: 2002 \apj 581, 1297

\bibitem[{{Juett} \& {Chakrabarty}(2006)}]{1916:juett06apj}
{Juett}, A.~M. \& {Chakrabarty}, D.: 2006 \apj 646, 493

\bibitem[{{Juett} et~al.(2006){Juett}, {Schulz} et~al.}]{juett06apj}
{Juett}, A.~M., {Schulz}, N.~S. et~al.: 2006 \apj 648, 1066

\bibitem[{{Kalberla} et~al.(2005){Kalberla}, {Burton} et~al.}]{kalberla05aa}
{Kalberla}, P.~M.~W., {Burton}, W.~B. et~al.: 2005 \aap 440, 775

\bibitem[{{Kallman} \& {Bautista}(2001)}]{kallman01apjs}
{Kallman}, T. \& {Bautista}, M.: 2001 \apjs 133, 221

\bibitem[{{Kallman} et~al.(2009){Kallman}, {Bautista}
  et~al.}]{1655:kallman09apj}
{Kallman}, T.~R., {Bautista}, M.~A. et~al.: 2009 \apj 701, 865

\bibitem[{{King} et~al.(2012){King}, {Miller} et~al.}]{1709:king12apj}
{King}, A.~L., {Miller}, J.~M. et~al.: 2012 \apjl 746, L20

\bibitem[{{King} et~al.(2013){King}, {Miller} et~al.}]{king13apj}
{King}, A.~L., {Miller}, J.~M. et~al.: 2013 \apj 762, 103

\bibitem[{{King} et~al.(2014){King}, {Walton} et~al.}]{1630:king14apjl}
{King}, A.~L., {Walton}, D.~J. et~al.: 2014 \apjl 784, L2

\bibitem[{{Kotani} et~al.(2000){Kotani}, {Ebisawa} et~al.}]{1915:kotani00apj}
{Kotani}, T., {Ebisawa}, K. et~al.: 2000 \apj 539, 413

\bibitem[{{Kubota} et~al.(2007){Kubota}, {Dotani} et~al.}]{1630:kubota07pasj}
{Kubota}, A., {Dotani}, T. et~al.: 2007 \pasj 59, 185

\bibitem[{{Lee} et~al.(2002){Lee}, {Reynolds} et~al.}]{1915:lee02apj}
{Lee}, J.~C., {Reynolds}, C.~S. et~al.: 2002 \apj 567, 1102

\bibitem[{{Liu} et~al.(2007){Liu}, {van Paradijs} et~al.}]{liu07aa}
{Liu}, Q.~Z., {van Paradijs}, J. et~al.: 2007 \aap 469, 807

\bibitem[{{Luketic} et~al.(2010){Luketic}, {Proga} et~al.}]{1655:luketic10apj}
{Luketic}, S., {Proga}, D. et~al.: 2010 \apj 719, 515

\bibitem[{{Madej} et~al.(2014){Madej}, {Jonker} et~al.}]{madej14mnras}
{Madej}, O.~K., {Jonker}, P.~G. et~al.: 2014 \mnras 438, 145

\bibitem[{{Martocchia} et~al.(2006){Martocchia}, {Matt}
  et~al.}]{1915:martocchia06aa}
{Martocchia}, A., {Matt}, G. et~al.: 2006 \aap 448, 677

\bibitem[{{Miller} et~al.(2002){Miller}, {Fabian} et~al.}]{1650:miller02apjl}
{Miller}, J.~M., {Fabian}, A.~C. et~al.: 2002 \apjl 570, L69

\bibitem[{{Miller} et~al.(2011){Miller}, {Maitra}
  et~al.}]{igrj1748:miller11apj}
{Miller}, J.~M., {Maitra}, D. et~al.: 2011 \apjl 731, L7

\bibitem[{{Miller} et~al.(2004){Miller}, {Raymond} et~al.}]{gx339:miller04apj}
{Miller}, J.~M., {Raymond}, J. et~al.: 2004 \apj 601, 450

\bibitem[{{Miller} et~al.(2006{\natexlab{a}}){Miller}, {Raymond}
  et~al.}]{1743:miller06apj}
{Miller}, J.~M., {Raymond}, J. et~al.: 2006{\natexlab{a}} \apj 646, 394

\bibitem[{{Miller} et~al.(2006{\natexlab{b}}){Miller}, {Raymond}
  et~al.}]{1655:miller06nat}
{Miller}, J.~M., {Raymond}, J. et~al.: 2006{\natexlab{b}} \nat 441, 953

\bibitem[{{Miller} et~al.(2008){Miller}, {Raymond} et~al.}]{1655:miller08apj}
{Miller}, J.~M., {Raymond}, J. et~al.: 2008 \apj 680, 1359

\bibitem[{{Miller} et~al.(2012){Miller}, {Raymond} et~al.}]{1743:miller12apjl}
{Miller}, J.~M., {Raymond}, J. et~al.: 2012 \apjl 759, L6

\bibitem[{{Miller} et~al.(2014){Miller}, {Raymond} et~al.}]{1305:miller14apj}
{Miller}, J.~M., {Raymond}, J. et~al.: 2014 \apj 788, 53

\bibitem[{{Mu{\~n}oz-Darias} et~al.(2008){Mu{\~n}oz-Darias}, {Casares}
  et~al.}]{gx339:munoz-darias08mnras}
{Mu{\~n}oz-Darias}, T., {Casares}, J. et~al.: 2008 \mnras 385, 2205

\bibitem[{{Mu{\~n}oz-Darias} et~al.(2009){Mu{\~n}oz-Darias}, {Casares}
  et~al.}]{0748:munoz09mnras}
{Mu{\~n}oz-Darias}, T., {Casares}, J. et~al.: 2009 \mnras L196

\bibitem[{{Neilsen} et~al.(2014){Neilsen}, {Coriat}
  et~al.}]{1630:neilsen14apjl}
{Neilsen}, J., {Coriat}, M. et~al.: 2014 \apjl 784, L5

\bibitem[{{Neilsen} \& {Homan}(2012)}]{1655:neilsen12apj}
{Neilsen}, J. \& {Homan}, J.: 2012 \apj 750, 27

\bibitem[{{Neilsen} \& {Lee}(2009)}]{1915:neilsen09nat}
{Neilsen}, J. \& {Lee}, J.~C.: 2009 \nat 458, 481

\bibitem[{{Neilsen} et~al.(2012){Neilsen}, {Petschek}
  et~al.}]{1915:neilsen12mnras}
{Neilsen}, J., {Petschek}, A.~J. et~al.: 2012 \mnras 421, 502

\bibitem[{{Neilsen} et~al.(2011){Neilsen}, {Remillard}
  et~al.}]{1915:neilsen11apj}
{Neilsen}, J., {Remillard}, R.~A. et~al.: 2011 \apj 737, 69

\bibitem[{{Netzer}(2006)}]{1655:netzer06apj}
{Netzer}, H.: 2006 \apjl 652, L117

\bibitem[{{Papitto} et~al.(2011){Papitto}, {D'A{\`i}}
  et~al.}]{igrj1748:papitto11aa}
{Papitto}, A., {D'A{\`i}}, A. et~al.: 2011 \aap 526, L3

\bibitem[{{Parmar} et~al.(2002){Parmar}, {Oosterbroek}
  et~al.}]{1624:parmar02aa}
{Parmar}, A.~N., {Oosterbroek}, T. et~al.: 2002 \aap 386, 910

\bibitem[{{Ponti} et~al.(2015){Ponti}, {Bianchi} et~al.}]{axj1745:ponti15mnras}
{Ponti}, G., {Bianchi}, S. et~al.: 2015 \mnras 446, 1536

\bibitem[{{Ponti} et~al.(2012){Ponti}, {Fender} et~al.}]{ponti12mnras}
{Ponti}, G., {Fender}, R.~P. et~al.: 2012 \mnras 422, L11

\bibitem[{{Ponti} et~al.(2014){Ponti}, {Mu{\~n}oz-Darias}
  et~al.}]{0748:ponti14mnras}
{Ponti}, G., {Mu{\~n}oz-Darias}, T. et~al.: 2014 \mnras 444, 1829

\bibitem[{{Proga} \& {Kallman}(2002)}]{proga02apj}
{Proga}, D. \& {Kallman}, T.~R.: 2002 \apj 565, 455

\bibitem[{{Rahoui} et~al.(2014){Rahoui}, {Coriat} et~al.}]{gx339:rahoui14mnras}
{Rahoui}, F., {Coriat}, M. et~al.: 2014 \mnras 442, 1610

\bibitem[{{Rao} \& {Vadawale}(2012)}]{1709:rao12apjl}
{Rao}, A. \& {Vadawale}, S.~V.: 2012 \apjl 757, L12

\bibitem[{{Sala} et~al.(2007){Sala}, {Greiner} et~al.}]{1655:sala07aa}
{Sala}, G., {Greiner}, J. et~al.: 2007 \aap 461, 1049

\bibitem[{{Schulz} \& {Brandt}(2002)}]{cirx1:schulz02apj}
{Schulz}, N.~S. \& {Brandt}, W.~N.: 2002 \apj 572, 971

\bibitem[{{Schulz} et~al.(2008){Schulz}, {Kallman} et~al.}]{cirx1:schulz08apj}
{Schulz}, N.~S., {Kallman}, T.~E. et~al.: 2008 \apj 672, 1091

\bibitem[{{Shakura} \& {Sunyaev}(1973)}]{shakura73aa}
{Shakura}, N.~I. \& {Sunyaev}, R.~A.: 1973 \aap 24, 337

\bibitem[{{Shidatsu} et~al.(2011){Shidatsu}, {Ueda}
  et~al.}]{gx339:shidatsu11pasj}
{Shidatsu}, M., {Ueda}, Y. et~al.: 2011 \pasj 63, 785

\bibitem[{{Shidatsu} et~al.(2013){Shidatsu}, {Ueda}
  et~al.}]{1305:shidatsu13apj}
{Shidatsu}, M., {Ueda}, Y. et~al.: 2013 \apj 779, 26

\bibitem[{{Shields} et~al.(1986){Shields}, {McKee} et~al.}]{shields86apj}
{Shields}, G.~A., {McKee}, C.~F. et~al.: 1986 \apj 306, 90

\bibitem[{{Sidoli} et~al.(2001){Sidoli}, {Oosterbroek}
  et~al.}]{1658:sidoli01aa}
{Sidoli}, L., {Oosterbroek}, T. et~al.: 2001 \aap 379, 540

\bibitem[{{Sidoli} et~al.(2002){Sidoli}, {Parmar} et~al.}]{gx13:sidoli02aa}
{Sidoli}, L., {Parmar}, A.~N. et~al.: 2002 \aap 385, 940

\bibitem[{{Stern} et~al.(2014){Stern}, {Behar} et~al.}]{stern14mnras}
{Stern}, J., {Behar}, E. et~al.: 2014 \mnras 445, 3011

\bibitem[{{Takahashi} et~al.(2008){Takahashi}, {Fukazawa}
  et~al.}]{1655:takahashi08pasj}
{Takahashi}, H., {Fukazawa}, Y. et~al.: 2008 \pasj 60, 69

\bibitem[{{Tarter} et~al.(1969){Tarter}, {Tucker} et~al.}]{tarter69apj}
{Tarter}, C.~B., {Tucker}, W.~H. et~al.: 1969 \apj 156, 943

\bibitem[{{Ueda} et~al.(2001){Ueda}, {Asai} et~al.}]{gx13:ueda01apjl}
{Ueda}, Y., {Asai}, K. et~al.: 2001 \apjl 556, L87

\bibitem[{{Ueda} et~al.(2010){Ueda}, {Honda} et~al.}]{1915:ueda10apj}
{Ueda}, Y., {Honda}, K. et~al.: 2010 \apj 713, 257

\bibitem[{{Ueda} et~al.(1998){Ueda}, {Inoue} et~al.}]{1655:ueda98apj}
{Ueda}, Y., {Inoue}, H. et~al.: 1998 \apj 492, 782

\bibitem[{{Ueda} et~al.(2004){Ueda}, {Murakami} et~al.}]{gx13:ueda04apj}
{Ueda}, Y., {Murakami}, H. et~al.: 2004 \apj 609, 325

\bibitem[{{Ueda} et~al.(2009){Ueda}, {Yamaoka} et~al.}]{1915:ueda09apj}
{Ueda}, Y., {Yamaoka}, K. et~al.: 2009 \apj 695, 888

\bibitem[{{van Peet} et~al.(2009){van Peet}, {Costantini}
  et~al.}]{0748:vanpeet09aa}
{van Peet}, J.~C.~A., {Costantini}, E. et~al.: 2009 \aap 497, 805

\bibitem[{{Wijnands} et~al.(2012){Wijnands}, {Yang}
  et~al.}]{1709:wijnands12mnras}
{Wijnands}, R., {Yang}, Y.~J. et~al.: 2012 \mnras 422, L91

\bibitem[{{Woods} et~al.(1996){Woods}, {Klein} et~al.}]{woods96apj}
{Woods}, D.~T., {Klein}, R.~I. et~al.: 1996 \apj 461, 767

\bibitem[{{Xiang} et~al.(2009){Xiang}, {Lee} et~al.}]{1624:xiang09apj}
{Xiang}, J., {Lee}, J.~C. et~al.: 2009 \apj 701, 984

\bibitem[{{Yamaoka} et~al.(2001){Yamaoka}, {Ueda} et~al.}]{1655:yamaoka01pasj}
{Yamaoka}, K., {Ueda}, Y. et~al.: 2001 \pasj 53, 179

\bibitem[{{Zhang} et~al.(2014){Zhang}, {Makishima} et~al.}]{1916:zhang14pasj}
{Zhang}, Z., {Makishima}, K. et~al.: 2014 \pasj 66, 120

\end{thebibliography}

\end{document}